# Coherent Modulation of Two-Dimensional Moiré States with On-Chip THz Waves

Yiliu Li[1,†], Eric A. Arsenault[1,*,†], Birui Yang[2], Xi Wang[3,4], Heonjoon Park[5], Yinjie Guo[2], Takashi Taniguchi[6], Kenji Watanabe[7], Daniel Gamelin[8], James C. Hone[9], Cory R. Dean[2], Sebastian F. Maehrlein[10], Xiaodong Xu[5,11], and Xiaoyang Zhu[1,‡]

[1] Department of Chemistry, Columbia University, New York, NY 10027, USA
[2] Department of Physics, Columbia University, New York, NY 10027, USA
[3] Department of Physics, Washington University, St. Louis, MO 63130, USA
[4] Institute of Materials Science & Engineering, Washington University, St. Louis, MO 63130, USA
[5] Department of Physics, University of Washington, Seattle, WA 98195, USA
[6] Research Center for Materials Nanoarchitectonics, National Institute for Materials Science, 1-1 Namiki, Tsukuba 305-0044, Japan
[7] Research Center for Electronic and Optical Materials, National Institute for Materials Science, 1-1 Namiki, Tsukuba 305-0044, Japan
[8] Department of Chemistry, University of Washington, Seattle, WA, USA
[9] Department of Mechanical Engineering, Columbia University, New York, NY 10027, USA
[10] Department of Physical Chemistry, Fritz Haber Institute of the Max Planck Society, Berlin, Germany
[11] Department of Materials Science and Engineering, University of Washington, Seattle, WA, USA

**ABSTRACT**. Van der Waals (vdW) structures of two-dimensional materials host a broad range of physical phenomena. New opportunities arise if different functional layers may be remotely modulated or coupled in a device structure. Here we demonstrate the in-situ coherent modulation of moiré excitons and correlated Mott insulators in transition metal dichalcogenide (TMD) homo- or hetero-bilayers with on-chip terahertz (THz) waves. Using common dual-gated device structures, each consisting of a TMD moiré bilayer sandwiched between two few-layer graphene (fl-Gr) gates with hexagonal boron nitride (h-BN) spacers, we launch coherent phonon wavepackets at ~0.4-1 THz from the fl-Gr gates by femtosecond laser excitation. The waves travel through the h-BN spacer, arrive at the TMD bilayer with precise timing, and coherently modulate the moiré excitons or the Mott states. These results demonstrate that the fl-Gr gates, often used for electrical control of the material properties, can serve as effective on-chip opto-elastic transducers to generate THz waves for the coherent control and vibrational entanglement of functional layers in commonly used moiré devices.

Two-dimensional (2D) vdW materials have emerged as a versatile platform for the exploration and realization of a broad range of physical phenomena, particularly quantum phenomena. These materials can be mechanically exfoliated down to the monolayer limit and artificially stacked with different sequences, twist angles, and symmetries. The tunable interactions between layers can dramatically change physical properties or introduce new ones. Exciting discoveries in vdW structures include, among others, correlated insulators (1–4), superconductors (5), tunable magnets (6), and both integer and fractional quantum anomalous Hall effects (7–9). Most studies on vdW structures to date have focused on emergent quantum phases of electrons and their static control by gate doping, electric fields, and magnetic fields. The dynamic counterparts of such controls, particularly by exciting phonon modes, can greatly expand our understanding of these quantum phenomena and the discovery of transient or hidden phases. Indeed, phonon excitations have been successfully demonstrated in transient control and engineering of novel physical properties in bulk crystalline solids, particularly quantum materials (10–14). However, extending the approaches to 2D moiré quantum matter is challenging, due to a mismatch between the wavelengths of infrared (IR) and THz radiation used to excite the phonon modes and the few micrometer sizes of typical vdW stacked devices. While it is possible to implement IR-THz excitation via sub-diffraction limit waveguides (14) or near-field scattering (15), these approaches require specific sample architectures and have not been attempted on moiré quantum matter.

In standard dual-gated moiré devices, the active moiré layers are encapsulated by h-BN dielectric spacers and fl-Gr top and bottom gates (1–5, 7–9, 16). For semiconducting TMDs, the moiré layers and the h-BN dielectric are optically transparent in the near-IR or longer wavelengths (in the linear excitation regime). Under these conditions, only the fl-Gr gates absorb light. Pulsed laser excitation can generate longitudinal acoustic waves and other coherent phonons (17), thus serving as sources for coherent waves. For 2D vdW materials, the dominant phonon modes in the low frequency region are the interfacial layer breathing ($\nu_{LB}$) and shear ($\nu_S$) modes in the hundreds of GHz to a few THz frequency range (18). These modes, which depend highly on interlayer interactions throughout a vdW stack, can be directly targeted by the coherent waves launched via fl-Gr gates. Thus, in addition to offering electrostatic control, the fl-Gr gates can serve as sources of tunable waves in the THz frequency range for the in-situ coherent control and coupling of moiré states in commonly used vdW devices.

We use the well-established mechanical exfoliation and transfer stacking techniques to prepare the fl-Gr/h-BN/TMD-BL/h-BN/fl-Gr vdW heterostructure (19) (Fig. 1a, Methods). For simplicity in data analysis, we use the same thickness ($d_{hBN}$) for the upper and the lower h-BN spacers. While the twist angle between the TMD monolayers is precisely determined, their orientations with respect to h-BN and fl-Gr are not specified. Fig. 1b shows a microscope image of the $\Delta\theta = 60\pm1°$ $WSe_2/WS_2$ device (D1, $d_{hBN} = 36\pm1$ nm). Spatially resolved reflectance mapping (Fig. 1c) integrated over the first moiré $WSe_2$ exciton features a homogeneous bilayer region highlighted with black dashed line. A representative reflectance spectrum is shown in Fig. 1d. As reported by Jin et al. (20), the formation of moiré excitons in the $WSe_2/WS_2$ heterobilayer is evidenced by the splitting of the single A-exciton $WSe_2$ monolayer peak into three peaks in the range of ~1.7-1.8 eV. These transitions are assigned to flat-bands due to localization of the $WSe_2$ A-excitons in periodic local potential wells in the moiré superlattice (20). For comparison, the $WS_2$ A-exciton at ~2.0 eV is less affected by the moiré superlattice and only shows broadening from that of the monolayer. Fig. S1-S4 provide characterization of the other samples investigated in this work.

We apply femtosecond pump-probe spectroscopy in the reflection geometry, as illustrated in Fig. 1a. The pump photon energy ($h\nu_1$ = 1.55 eV) is below the optical bandgap of the TMD layers and only excites the top and bottom fl-Gr gates. A probe pulse with photon energy ($h\nu_2$) covering the first $WSe_2$ moiré exciton resonance is used to detect the coherent oscillations in the WSe$_2$/WS$_2$ heterobilayer or the WSe$_2$/WSe$_2$ homobilayer region. The fl-Gr serves as an opto-elastic transducer and the dynamical strain field generates a broadband coherent phonon wavepacket, which propagates through the h-BN spacer and reaches the TMD-BL at a time delay determined by the spacer thickness and phonon group velocity. Figs. 2a-c show transient reflectance ($\Delta R/R$) spectra as a function of probe energy and pump-probe delay ($\Delta t$) from three angle-aligned (($\Delta\theta = 60\pm1°$, $\Delta\theta = 0\pm1°$) WSe$_2$/WS$_2$ devices, with h-BN thickness of $d_{hBN}$ = 36±1 (D1), $d_{hBN}$ = 18±1 (D2) and 36±1 nm (D3), respectively. The key feature of each spectrum is the presence of highly oscillatory dynamics that are more clearly resolved via the second derivative (with respect to Δt), shown in Fig. 2d-f. For D2 and D3, a π-phase shift in the coherent oscillations of the main moiré exciton peak at 1.683 eV can be clearly observed, as expected for an optical transition modulated by a coherent mode (21). This π-phase shift for D1 likely appears outside of the employed probe window as the main moiré exciton is blue shifted in this sample. Moreover, for all samples, the arrival of the coherent oscillation signal at the TMD sensing layer is delayed, with the delay time scaling with the thickness of the h-BN spacer.

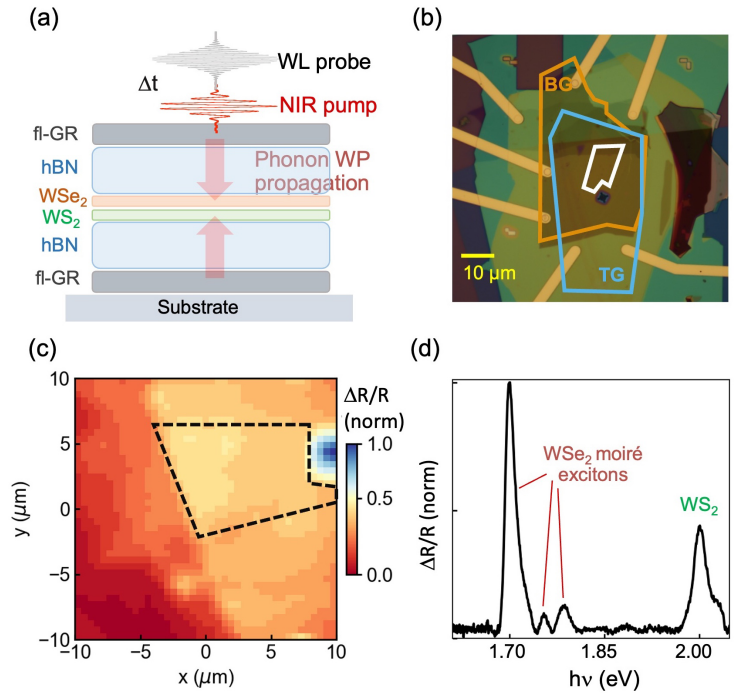

**Figure 1. The vdW structure and pump-probe experiment.** (a) Schematic of the multilayer vdW structure and illustration of the experiment. WL: white light, NIR: near infrared, WP: wavepacket, fr-Gr: few-layer graphene; (b) Optical image of the device (D1, $\Delta\theta = 60\pm1°$); TG and BG: top (blue outline) and bottom (orange outline) gates of fl-Gr; the $WSe_2/WS_2$ heterobilayer region is shown by white outline. (c) Integrated reflectance contrast map of the device. The black dashed line highlights the $WSe_2/WS_2$ heterobilayer region. (d) The steady-state reflectance spectrum of the structure showing the $WSe_2$ moiré excitons ~1.7-1.8 eV and the $WS_2$ exciton at ~2.0 eV. All spectra above were collected at sample temperature of 11 K.

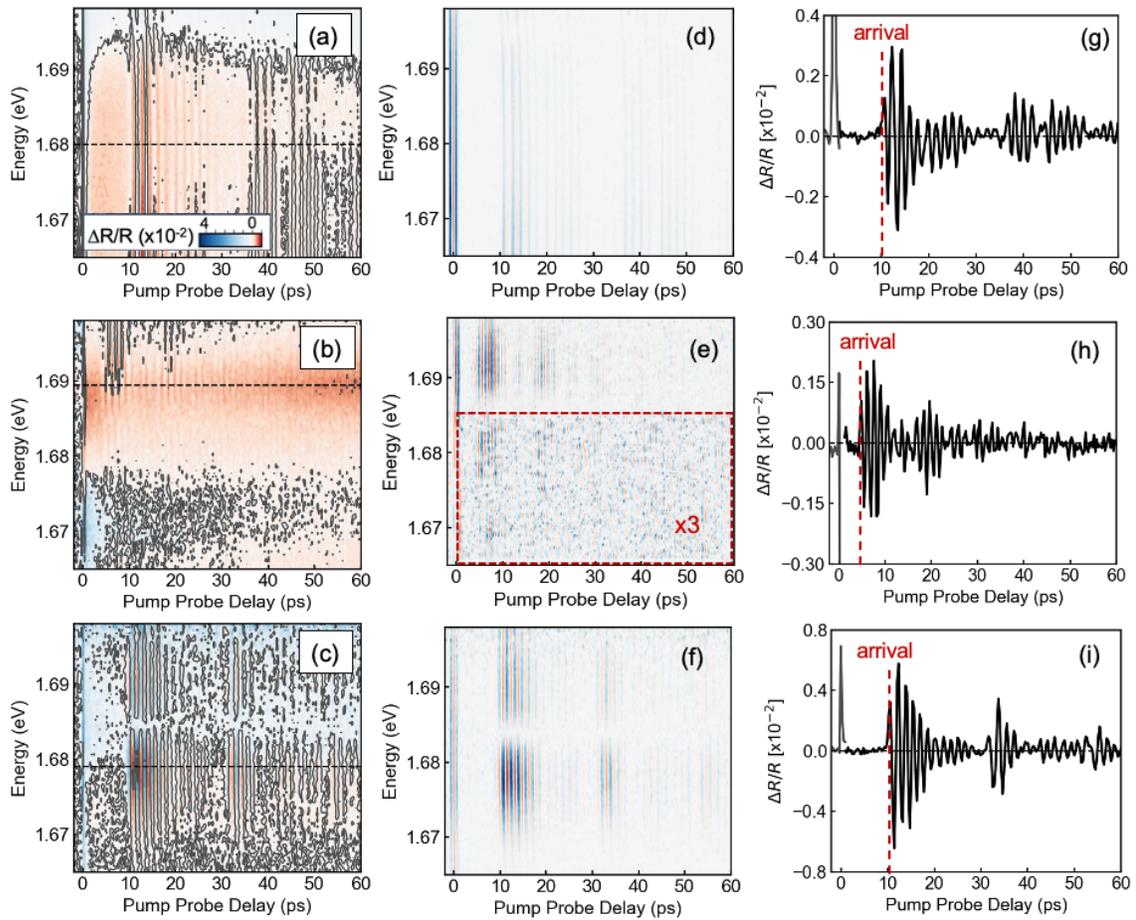

**Figure 2. Time domain spectra of coherent phonon transport.** Transient reflectance spectra plotted as a function of probe energy (spanning the first moiré exciton of WSe$_2$) of WSe$_2$/WS$_2$ devices with $\Delta\theta$ = 60±1° (a) $d_{hBN}$ = 36±1 nm (D1, the pump fluence employed for the transient measurements was 13.7 μJ/cm$^2$, sample temperature was 8 K) and $\Delta\theta$ = 0±1° (b) $d_{hBN}$ = 18±1 nm (D2, the pump fluence employed for the transient measurements was 84.1 μJ/cm$^2$, sample temperature was 11 K) and (c) $d_{hBN}$ = 36±1 nm (D3, the pump fluence employed for the transient measurements was 13.7 μJ/cm$^2$, sample temperature was 8 K), respectively. (d) - (f) Second derivatives of spectra in (a) - (c), respectively. In (e), the signal intensity has been multiplied by a factor of three in the red-dashed box. (g) - (i) Line cuts from (a) - (c) showing background-subtracted transient reflectance at fixed probe photon energies, as indicated by the dashed lines in (a) - (c). The delayed arrivals of the oscillations are labeled on the plot.

To analyze the coherent oscillations, we take line cuts of the ΔR/R spectra at fixed probe photon energies (dashed lines in Figs. 2a-c), with the slow-varying background/offset subtracted. The results are shown in Figs. 2g-i, for devices D1, D2 and D3, respectively. The time traces show oscillations nearly symmetric about zero. We can clearly observe onset time delays of $t_{delay}$ = 10.3±0.6 ps, 4.6±0.3 ps and 10.1±0.4 ps for $d_{hBN}$ = 36±1 nm, 18±1 nm, and 36±1 nm, respectively, indicating the arrivals of the phonon wavepackets at the TMD moiré bilayer and the initiation of coherent modulation of the moiré excitons. The onset time delay is related to the group velocity ($v_{ph}^g$) of the phonon wavepacket by

$$t_{delay} = d_{hBN} \cdot \left( \frac{1}{v_{ph}^g} - \frac{n}{C} \right) \quad (1),$$

where $n$ is the refractive index and $C$ the speed of the probe light pulse; since $C$ is many-orders of magnitude larger than $v_{ph}^g$, the second term in the parentheses can be neglected. This gives the phonon group velocity of $v_{ph}^g \approx d_{hBN}/t_{delay}$ = 3.7±0.2 km/s, in agreement with the longitudinal acoustic phonon velocity in h-BN (22).

The above findings are repeated in an additional device with similar structure: an AB stacked $WSe_2/WSe_2$ device with 2.7° twist angle (D4) with top and bottom fl-Gr electrodes and h-BN spacer thickness of 39±1 nm (Fig. S3). In a control experiment, we fabricated a $WSe_2/WS_2$ heterobilayer with h-BN encapsulation, but without the top and bottom fl-Gr layers. Under the same experimental conditions as those in Fig. 2 and Fig. S3, we observe no oscillations in transient reflectance spectra (Fig. S4); this is expected from the absence of the fl-Gr opto-elastic transducer.

Having established the effective launching of phonon wavepackets in the fl-Gr, we now address the phonon modes involved. Figs. 3a-d show the Fourier transform (FT) of the time-domain coherent phonon spectra from the four devices (Figs. 2g-2i, and Fig. S3e.) Each FT spectrum (colored curve) features well-resolved peaks in the frequency range ≤ 1 THz. In the ≤ 1THz region, the phonon spectra of 2D layered vdW systems are composed of interfacial layer breathing ($\nu_{LB}$; out-of-plane) and shear ($\nu_S$; in-plane) modes (23–25) that modulate interlayer coupling between the TMDs, and thus, the moiré exciton transitions. The material composition, layer number, and relative twist angles between layers all directly influence the interfacial mode density and spectral structure as is well-captured by the range of samples studied in this work.

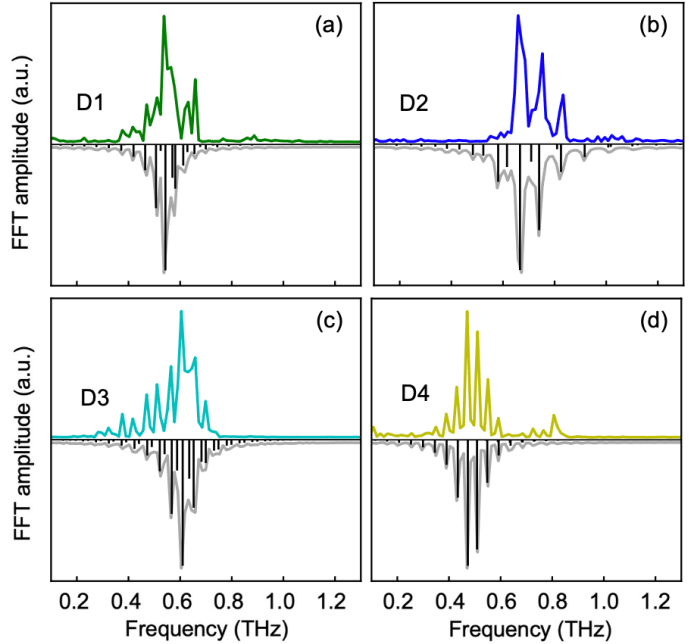

**Figure 3. Frequency domain data: experiments and calculation from linear chain model.** (a)-(d) Experimental FFT spectra for samples D1, D2, D3, and D4, respectively, shown as positive signal in colors. Fits to the LCM are shown as negative signal in each panel. The stick spectra are shown in black and the spectra with (Lorentzian) broadening are shown in gray. All fit parameters are provided in Table S1.

For multilayer vDW materials, the observed mode structure can be described by a one-dimensional linear chain model (LCM, Fig. S6) as interlayer interactions are much weaker than intralayer interactions (23, 26). Representing each layer as an effective mass per unit area, $m$, the experimentally observed low-frequency phonon spectra can be parametrized by the layer number, $N$, of each material and the force constants per unit area, $\alpha$, of each interlayer interaction. As shown by Lin et al., relative intensities can also be extracted via an eigenvector projection of the multilayer phonons onto the bilayer phonons (26). In our work, the double h-BN encapsulation of the TMD bilayer can lead to very closely spaced transitions — particularly for asymmetric interlayer coupling or thickness between the top and bottom h-BN — so we have further included a broadening parameter to account for the finite frequency resolution of our experiments (~13 GHz) and lifetime broadening of the phonon modes. Further details on the LCM can be found in the SI.

Fig. 3 contains the LCM fits (inverted on the y-axis) for the four samples (stick spectra shown in black and broadened spectra shown in gray). We find that the LCM prediction for the $\nu_{LB}$ modes of the h-BN encapsulated TMD bilayer accounts for a majority of the observed mode structure. In contrast, the $v_S$ mode LCM prediction fails to capture the correct mode spacing based on the known thickness range of the encapsulating h-BN (Fig. S7). From the LCM simulation, we find that the force constants for the $v_{LB}$ mode between a h-BN layer and a TMD layer is 15~35% of the values for the like-kind interfaces, h-BN/h-BN or TMD/TMD. The weak interlayer interactions between the h-BN and the TMDs are expected from atomic and geometric mismatch; the latter results from the fact the h-BN was not angle-aligned with the TMD bilayer during the stacking process. The overall shape of the phonon spectrum depends on the top versus bottom parameters (including bilayer composition, interlayer force constants, and h-BN layer number).

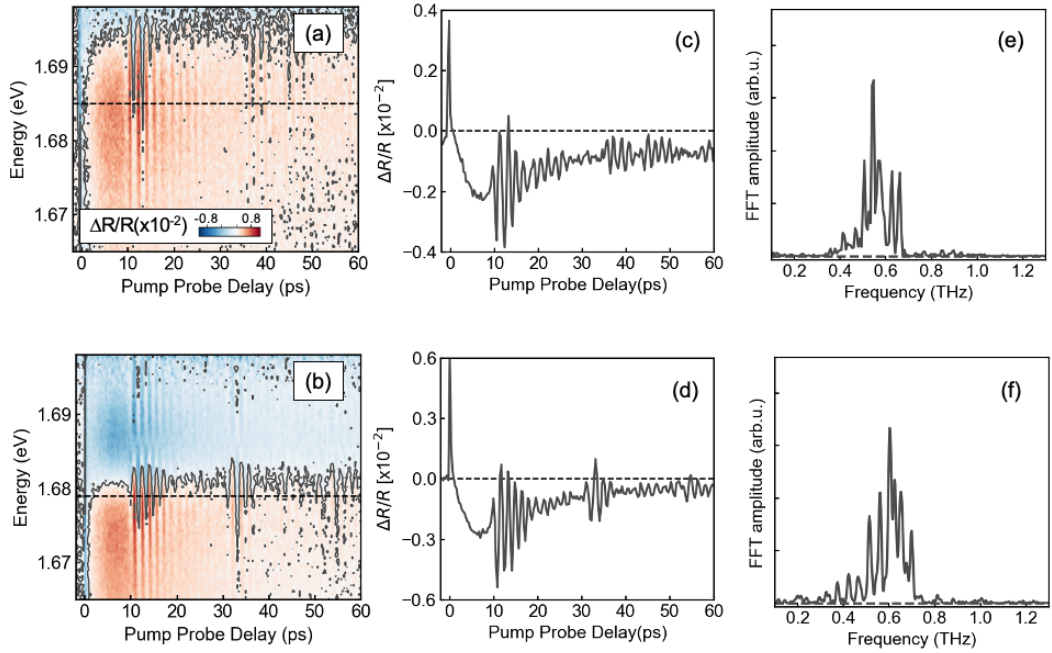

**Figure 4. Coherent modulation of Mott states.** Transient reflectance spectra for the $v = -1$ state plotted as a function of probe energy (spanning the first moiré exciton of $WSe_2$) of $WSe_2/WS_2$ devices with (a) $\Delta\theta = 60\pm1°$ (D1, the pump fluence employed for the transient measurements was 13.7 μJ/cm$^2$, sample temperature was 11 K) and (b) $\Delta\theta = 0\pm1°$ (D3, the pump fluence employed for the transient measurements was 13.7 μJ/cm$^2$, sample temperature was 8 K). (c), (d) Slice at a fixed probe energy corresponding to the dashed line in (a) and (b), respectively. (e), (f) Experimental FFT spectra for the $v = -1$ state for D1 and D3, respectively, following population subtraction.

The coherent modulations of the $WSe_2/WS_2$ heterobilayer are also observed when the active moiré structure is doped to form Mott insulator states (27). We demonstrate this for D1 and D3 which host one- and two-hole ($v = -1, -2$), and one- and two-electron ($v = 1, 2$) correlated insulator states, in addition to the charge neutral state ($v = 0$). These states are seen in the steady state reflectance spectrum in the moiré exciton region as a function of gate voltage ($V_g$) applied to the fl-Gr gates to dope the WSe$_2$/WS$_2$ heterobilayer (Figs. S5a-c) (28, 29). A reduction in the effective dielectric constant upon formation of correlated insulator states allows them to be revealed in the $V_g$-dependent reflectance spectrum via exciton sensing (2, 28, 30–32). We note that in the transient measurements, we focus on hole doping because we monitor the moiré excitons localized mainly in WSe$_2$ where the holes reside due to type II band alignment (27). Figs. 4a and 4b present pump-probe spectra for the $v = -1$ Mott state of D1 and D3, respectively. The corresponding data for the $v = -2$ state of D1 can be found in Fig. S5d-f.

Similar to the data from the undoped devices (Figs. 2 and 3), we observe coherent modulations in the transient reflectance spectra of the doped states. Note that unlike the undoped states, additional decay/recovery dynamics are observed (Fig. 4b and 4c) and attributed to disordering/re-ordering due to electronic excitation of the Mott states, as detailed elsewhere (27). Frequency analysis of the doped (D1: Fig. 4e, D3: Fig. 4f) versus undoped (D1: Fig. 3a, D3: Fig. 3c) states show very similar patterns, demonstrating that the same phonon wavepacket launched from the fl-Gr gates impinges on the Mott states. The coherent modulation of the electronic structure of the WSe$_2$/WS$_2$ moiré bilayer, as revealed by the moiré excitons in Fig. 2, is expected for all electronic properties — including the Mott states in Fig. 4. However, in the present study based on optical reflectance spectra, we are not able to strictly deconvolute the modulations of the Mott states from that of the moiré excitons. To this end, future experiments employing orthogonal probe schemes, e.g., transient THz conductivity (33), are warranted to directly isolate the modulation in the Mott states.

**Concluding Comments.** Using common device structures of TMD moiré bilayers sandwiched between fl-Gr gates with h-BN spacers, we demonstrate the on-chip generation of THz wavepackets from the fl-Gr gates and realize coherent modulation of the moiré states in the TMD bilayers by these wavepackets. The frequency range of the phonon wavepackets launched from each fl-Gr is a strong function of its thickness, which has not yet been fully explored here. While the 7±1 to 13±1 nm fl-Gr electrodes in this work can launch wavepackets which fully cover the interlayer modes observed from ~0.4-1.0 THz, this frequency range can be increased well into the multi-THz range as the fl-Gr thickness is decreased to the bilayer limit (34). Moreover, varying the relative thicknesses of the two h-BN spacers or the relative thickness of the two fl-Gr gates can create broadly tunable interference patterns, which is another tuning knob in the coherent control of the moiré states. Such tunability, along with differing material and twist angles, as well as the ability of the LCM to predict such spectra accurately, provides an effective toolset of on-chip THz wave generation for the coherent control and modulation of moiré states in vdW heterostructures.

**Acknowledgements.** XYZ acknowledges support by DOE-BES under award DE-SC0024343. XDX acknowledges support for sample fabrication and characterization as part of Programmable Quantum Materials, an Energy Frontier Research Center funded by the U.S. Department of Energy (DOE), Office of Science, Basic Energy Sciences (BES), under award DE-SC0019443. CRD and JCH acknowledge support for sample fabrication by the Materials Science and Engineering Research Center (MRSEC) through NSF grant DMR-2011738. XYZ acknowledges support for simulation and analysis by the Air Force Office of Scientific Research under award number FA9550-22-1-0389. EAA gratefully acknowledges support from the Simons Foundation as a Junior Fellow in the Simons Society of Fellows (965526). K.W. and T.T. acknowledge support from the JSPS KAKENHI (Grant Numbers 21H05233 and 23H02052) and World Premier International Research Center Initiative (WPI), MEXT, Japan for the growth of h-BN crystals.

**Author Contributions.** YL, EAA, and XYZ conceived this work. YL and EAA carried out all spectroscopic measurements. EAA performed the linear chain model calculation with inputs from SM. BY was responsible for sample (D1 and D4) fabrication and characterization, under the supervision of CRD and JH. HP was responsible for sample (D2) fabrication and characterization, under the supervision of XX. YL was responsible for sample (D5) fabrication and characterization, under the supervision of XYZ. TT and KW provided the h-BN crystal. YG performed sample mounting. The manuscript was prepared by YL, EAA and XYZ in consultation with all other authors. XYZ supervised the project. All authors read and commented on the manuscript.

**Competing Interests**. The authors declare no competing interests.

**Data Availability Statement**. The data that support the plots within this paper are available from the corresponding authors upon reasonable request.

SUPPLEMENRTARY INFORMATION

**Coherent Modulation of Two-Dimensional Moiré States with On-Chip THz Waves**

Yiliu Li[1,†], Eric A. Arsenault[1,*,†], Birui Yang[2], Xi Wang[3,4], Heonjoon Park[5], Yinjie Guo[2], Takashi Taniguchi[6], Kenji Watanabe[7], Daniel Gamelin[8], James C. Hone[9], Cory R. Dean[2], Sebastian F. Maehrlein[10], Xiaodong Xu[5,11], and Xiaoyang Zhu[1,‡]

[1] Department of Chemistry, Columbia University, New York, NY 10027, USA
[2] Department of Physics, Columbia University, New York, NY 10027, USA
[3] Department of Physics, Washington University, St. Louis, MO 63130, USA
[4] Institute of Materials Science & Engineering, Washington University, St. Louis, MO 63130, USA
[5] Department of Physics, University of Washington, Seattle, WA 98195, USA
[6] Research Center for Materials Nanoarchitectonics, National Institute for Materials Science, 1-1 Namiki, Tsukuba 305-0044, Japan
[7] Research Center for Electronic and Optical Materials, National Institute for Materials Science, 1-1 Namiki, Tsukuba 305-0044, Japan
[8] Department of Chemistry, University of Washington, Seattle, WA, USA
[9] Department of Mechanical Engineering, Columbia University, New York, NY 10027, USA
[10] Department of Physical Chemistry, Fritz Haber Institute of the Max Planck Society, Berlin, Germany
[11] Department of Materials Science and Engineering, University of Washington, Seattle, WA, USA

---

[†]These authors contributed equally.
[*] Junior Fellow in the Simons Society of Fellows, Simons Foundation.
[‡] To whom correspondence should be addressed. E-mail: xyzhu@columbia.edu.

**Methods**

**Device Fabrication**

**Device 1 (60° WSe₂/WS₂)** The WSe$_2$ and WS$_2$ monolayers are mechanically exfoliated from lab grown bulk WSe$_2$ crystals and commercially purchased bulk WS$_2$ crystals from HQ Graphene, respectively. Prior to transfer, the crystal orientation of WSe$_2$ and WS$_2$ monolayers are first determined by polarization resolved second harmonic generation (P-SHG) measurement. Then the monolayers are stacked together using dry-transfer technique with a polycarbonate (PC) stamp. To distinguish between R-stack and H-stack samples, P-SHG measurement is applied again to the heterobilayer region after the device is fabricated, with results compared to individual monolayers. The error bars for the angle determination are well within $\pm 1°$; the same applies to Devices 2, 3 and 5 below. The samples are grounded via graphite contacts connected to the heterobilayers, and single crystal hBN dielectrics and graphite gates are used to encapsulate the device and provide control to the charrier density and displacement field. SF6 radiofrequency plasma is applied to the stack to etch the encapsulating hBN and create connection to the graphite gates contacts. All electrodes are defined with electron beam lithography and made of a three-layer metal film of Cr/Pd/Au (3 nm/17 nm/60 nm).

**Device 2 (0° WSe₂/WS₂)** The complete sample fabrication details can be found in Ref. (*2*). In brief, platinum contacts are prefabricated using conventional e-beam lithography, on a bottom gate with a stack of hBN/graphite. Transition metal dichalcogenide (WS$_2$ and WSe$_2$) monolayers are mechanically exfoliated. Polycarbonate/polydimethylsiloxane stamps are used to pick up the WS$_2$ and WSe$_2$ flakes sequentially and melted down on the bottom gate. The crystal orientation is determined by second harmonic generation for the heterostructure alignment during fabrication. To clean the heterobilayer, AFM flattening is performed with a Bruker Dimension Icon AFM(*3*). The twist angle and moiré period are confirmed through PFM as mentioned elsewhere(*1*). In the end, the top graphite-hBN stack is transferred onto the sample to finish the dual gated geometry.

**Device 3 (0° WSe₂/WS₂)** h-BN, graphite, and transition metal dichalcogenide monolayers (WS$_2$, WSe$_2$) are mechanically exfoliated onto SiO$_2$/Si substrates. The WS$_2$ and WSe$_2$ monolayers are characterized by SHG to check crystal orientation. A standard polycarbonate-based dry transfer process is used to create a bottom gate structure by picking up hBN and graphite sequentially and melting down on a SiO$_2$/Si substrate. Conventional e-beam lithography is used to create platinum contacts and gold pads on the prefabricated bottom gate. The WS$_2$ and WSe$_2$ flakes are picked up with the polymer stamp and melted down on the bottom gate. A Bruker Dimension Icon atomic force microscope (AFM) is used in contact mode to clean and push out the bubbles in the heterobilayer. Piezoresponse force microscopy (PFM) is also performed to confirm the moiré wavelength and twist angle as mentioned elsewhere(*1*). Finally, the top gate structure composed of a graphite-hBN stack is transferred onto the sample to complete the dual gate geometry.

**Device 4 (60° WSe$_2$/WSe$_2$)** The WSe$_2$ monolayer are mechanically exfoliated from lab grown bulk WSe$_2$ crystals. The monolayers are stacked together using dry-transfer technique with a polycarbonate (PC) stamp. The twist homobilayer region is stacked using the tear-and-stack technique. Piezoresponse force microscopy (PFM) is performed to confirm the moiré wavelength and twist angle as mentioned elsewhere(*1*) The samples are grounded via graphite contacts connected to the homobilayers, and single crystal hBN dielectrics and graphite gates are used to encapsulate the device and provide control to the charrier density and displacement field. All electrodes are defined with electron beam lithography and made of a three-layer metal film of Cr/Pd/Au (3 nm/17 nm/60 nm).

**Device 5 (control device 0° WSe$_2$/WS$_2$)** The WSe$_2$ and WS$_2$ monolayers are mechanically exfoliated from lab grown bulk WSe$_2$ crystals and commercially purchased bulk WS$_2$ crystals from HQ Graphene, respectively. Prior to transfer, the crystal orientation of WSe$_2$ and WS$_2$ monolayers are first determined by polarization resolved second harmonic generation (P-SHG) measurement. Then the monolayers are stacked together using dry-transfer technique with a polycarbonate (PC) stamp. To distinguish between R-stack and H-stack samples, P-SHG measurement is applied again to the heterobilayer region after the device is fabricated, with results compared to individual monolayers.

**Spectroscopic Measurements**

For the spectroscopic measurements, the sample is cooled to the desired temperature (6-10 K) under vacuum (<10$^{-6}$ torr) with a closed-cycle liquid helium cryostat (Fusion X-Plane, Montana Instruments). Steady state reflectance measurements are carried out using a 3200 K halogen lamp (KLS EKE/AL). Following collimation, the light is directed onto the sample/substrate with a 100X, 0.75 NA objective. The reflected light is collected by the same objective and then dispersed with a spectrometer onto an InGaAs array (PyLoN-IR, Princeton Instruments). The steady state reflectance spectrum is obtained by contrasting the reflected signal from the sample ($R$) and substrate ($R_0$) as follows: $(R-R_0)/R_0$. For the spatially resolved reflectance experiments, a dual-axis galvo mirror scanning system is employed(*4*). Following spatial scanning of the sample as controlled via the angles of the galvo mirrors, the reflected light is spatially filtered through a pinhole and collected by the detector.

The pump-probe experiments (described in detail in Refs.(*5, 6*)) are seeded by femtosecond pulses (250 kHz, 1.55 eV, 100 fs) generated by a Ti:sapphire oscillator (Mira 900, Coherent) and regenerative amplifier (RegA9050, Coherent). The output is then split to form the pump and probe arms. For the probe, a fraction of the fundamental is focused into a sapphire crystal to generate a white light continuum which is then spectrally filtered (750±40 nm bandpass filter for D1, D3, D4, or 700 nm long pass and 750nm short pass filters for D2) to cover the lowest energy WSe$_2$ moiré exciton. The pump beam (800 nm, fundamental) is then directed towards a motorized delay stage to control the time delay, $\Delta t$, and passed through an optical chopper to generate pump-on and -off signals. Following this, the pump and probe arms are directed collinearly to the sample through an objective (100X, 0.75 NA). The same objective is used to collect the reflected light which is spectrally filtered to remove the pump and dispersed onto an InGaAs detector array (PyLoN-IR, Princeton Instruments). The pump-on and -off spectra at varying $\Delta t$ are then used to calculate the transient reflectance signal ($\Delta R/R$).

**Data Analysis**

In the following we describe the data processing involved in analyzing both the overall transient signal and the frequency domain data. The data is first collected as a function of pump-probe delay. To analyze the time traces, the signal in a ~0.01 eV window about the maximum of the lowest energy moiré band of WSe$_2$ is averaged. To isolate the coherent oscillations from the abovementioned time traces, fits using a biexponential function (with terms of the form $A_i \exp[t/\tau_i]$) are performed and subtracted from the overall time trace. The resulting coherent oscillations are then subjected to a Fourier transform to yield the frequency domain response. Additionally, the first ~500 fs are disregarded from all fits to suppress the involvement of artifacts from the pulse overlap region. Based on the time step size of 300 fs and the number of pump probe delay steps, $N$, used in the Fourier transform, the experimental frequency resolution can be estimated to be ~13 GHz.

**Linear Chain Model**

The system for the LCM is taken to be the vDW stack consisting of top and bottom hBN in addition to the TMD bilayer. Here, we assume that the coupling of the TMD bilayer to the top and bottom graphene is negligibly small. Each layer is then approximated as a ball with corresponding mass per unit area, $m$, calculated from known experimental bulk parameters, Table 1 from Ref. (*7*). Only nearest neighbor interactions between layers $i$ and $j$ are allowed and are given by the force constant $\alpha_{ij}$. The force constant characterizing hBN-hBN interactions (either for layer breathing (LB) or shear (S) modes), $\alpha_{\text{hBN-hBN}}$, is obtained from Ref. (*7*). All other force constants ($\alpha_{\text{hBN-WSe2}}$, $\alpha_{\text{WSe2-WS2}}$, and $\alpha_{\text{hBN-WS2}}$) are taken to be fit parameters. The LCM frequencies, $\omega$, are found by solving the corresponding system of $N_{\text{tot}}$ (total layer number) equations given by (*8*):

$$\omega_i^2 \mathbf{M} \mathbf{u}_i = \mathbf{D} \mathbf{u}_i$$

in matrix form where $\mathbf{u}_i$ is the eigenvector of the $i^{\text{th}}$ phonon with frequency $\omega_i$, $\mathbf{M}$ is the mass matrix, and $\mathbf{D}$ is the tridiagonal (i.e., nearest neighbor) $N_{\text{tot}}$ by $N_{\text{tot}}$ force constant matrix for either the LB or S modes. As demonstrated by Lin et al., an eigenvalue projection method can be further implemented to estimate the expected intensity of mode $i$, $I_i$, of the extended vDW multilayer stack (ML) versus the unencapsulated bilayer (BL) which is given by (*8*):

$$I_i = \left| \langle \mathbf{u}_{\text{BL}} | \mathbf{u}_{\text{ML},i} \rangle \right|^2$$

where $\mathbf{u}_{\text{BL}}$ is the eigenvector of the original mode in the BL stack and $\mathbf{u}_{\text{ML},i}$ is the eigenvector of the $i^{\text{th}}$ mode in the ML stack.

To reproduce the observed phonon spectra in our work, we perform a fit of the data to the LCM parametrized by the three unknown force constants listed above, the layer number of top and bottom hBN ($N_{\text{hBN (top)}}$ and $N_{\text{hBN (bot.)}}$), and an additional Lorentzian broadening parameter which is incorporated to account for the finite frequency resolution. We note that the hBN layer numbers are heavily constrained parameters based on the hBN thicknesses measured via AFM. Briefly, the dominant contribution of each parameter is as follows: $\alpha_{\text{WSe2-WS2}}$ controls the average central frequency and intensity of the observed modes; $\alpha_{\text{hBN-WSe2}}$ and $\alpha_{\text{hBN-WS2}}$ control the intensity of the observed modes; $N_{hBN\,(top)}$ and $N_{hBN\,(bot.)}$ control the total mode number of modes; and the broadening parameter controls the finite frequency resolution of the LCM fit. While the above force constants do have small effects over the mode spacing, $\alpha_{hBN\text{-}hBN}$ is seen to completely dominate this spacing and correspondingly the overall mode density (Fig. S7). Based on the known values of $\alpha_{hBN\text{-}hBN}$ for the LB and S modes, $9.83\times10^{19}$ Nm$^{-3}$ and $1.83\times10^{19}$ Nm$^{-3}$, respectively, the observed mode spacing and density in all samples can only be reliably reproduced via the LCM for the LB modes (*7*). Therefore, all fits correspond to LB modes unless otherwise specified.

Table S1. LCM Fit Parameters for D1-D4

|    | $N_{hBN\,(top)}$ (layer number) | $N_{hBN\,(bot.)}$ (layer number) | $\alpha_{hBN\text{-}WSe2}$ (Nm$^{-3}$) | $\alpha_{WSe2\text{-}WS2}$ (Nm$^{-3}$) | $\alpha_{hBN\text{-}WS2}$ (Nm$^{-3}$) | FWHM (Hz) |
|----|---|---|---|---|---|---|
| D1 | 110 | 123 | $1.5\times10^{19}$ | $2.45\times10^{19}$ | $1.7\times10^{19}$ | $4.5\times10^{9}$ |
| D2 | 51 | 57 | $3.4\times10^{19}$ | $3.9\times10^{19}$ | $3\times10^{19}$ | $1.05\times10^{10}$ |
| D3 | 108 | 110 | $2.85\times10^{19}$ | $3.1\times10^{19}$ | $1.9\times10^{19}$ | $4.5\times10^{9}$ |
| D4 | 118 | 117 | $2\times10^{19}$ | $2.2\times10^{19}$ | $1.55\times10^{19}$ | $3\times10^{9}$ |

**Methods References**

**Extended Data**

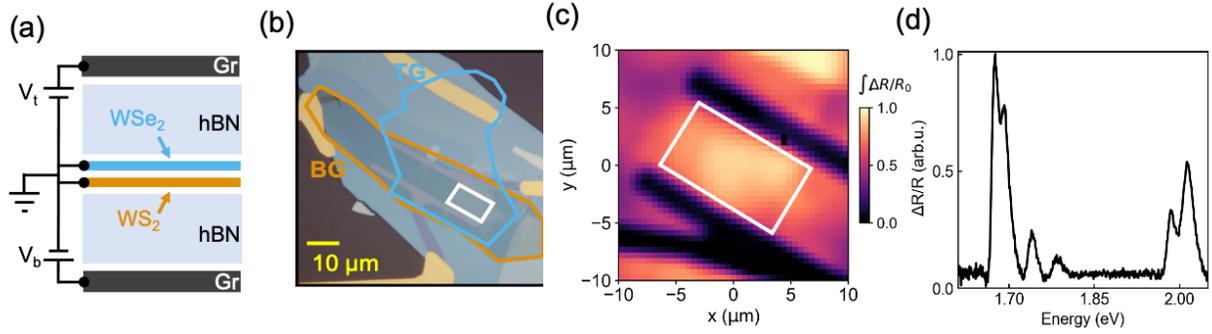

**Figure S1. WSe$_2$/WS$_2$ device D2 ($\theta = 0\pm1°$): structure and characterization.** (a) Schematic of dual-gated device (D2) showing the top ($V_t$) and bottom ($V_b$) gates consisting of few-layer graphite (Gr) and hexagonal boron nitride (hBN) on either side of the WSe$_2$/WS$_2$ heterobilayer. (b) Optical image of the device where the top and bottom gates are labeled, along with outlines showing the Gr electrodes (blue and orange lines corresponding to the top and bottom electrodes, respectively) and the WSe$_2$/WS$_2$ overlap region (white line). (c) Integrated steady-state reflectance contrast mapping of the device. The white line approximately highlights the WSe$_2$/WS$_2$ overlap region. (d) Steady-state reflectance spectrum of D2. All spectra were collected at sample temperature of 8 K.

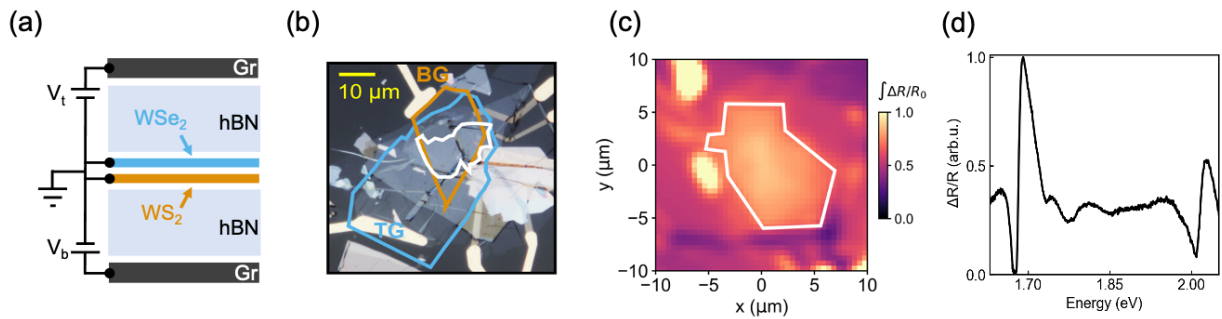

**Figure S2. WSe$_2$/WS$_2$ device D3 ($\theta = 0\pm1°$): structure and characterization.** (a) Schematic of dual-gated device (D3) showing the top ($V_t$) and bottom ($V_b$) gates consisting of few-layer graphite (Gr) and hexagonal boron nitride (hBN) on either side of the WSe$_2$/WS$_2$ heterobilayer. (b) Optical image of the device where the top and bottom gates are labeled, along with outlines showing the Gr electrodes (blue and orange lines corresponding to the top and bottom electrodes, respectively) and the WSe$_2$/WS$_2$ overlap region (white line). (c) Integrated steady-state reflectance contrast mapping of the device. The white line approximately highlights the WSe$_2$/WS$_2$ overlap region. (d) Steady-state reflectance spectrum of D3. All spectra were collected at sample temperature of 8 K.

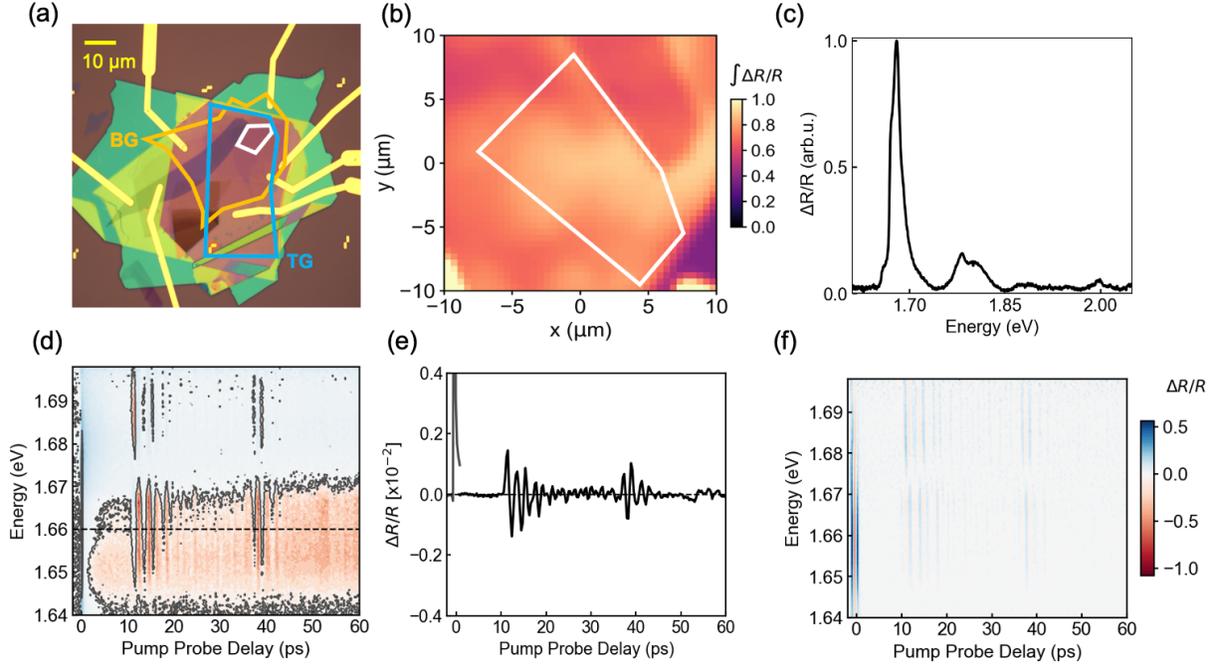

**Figure S3. WSe$_2$/WSe$_2$ device D4 ($\theta$ = 2.7±1°): structure and characterization.** (a) Optical image of the device where the top and bottom gates are labeled, along with outlines showing the Gr electrodes (blue and orange lines corresponding to the top and bottom electrodes, respectively) (b) Integrated steady-state reflectance contrast mapping of the device. The white line approximately highlights the WSe$_2$/WSe$_2$ overlap region. (c) Steady-state reflectance spectrum of D4. (d) Transient reflectance spectra plotted as a function of probe energy (covering the first moiré exciton of WSe$_2$). The pump fluence employed for the transient measurements was 84.1 µJ/cm$^2$. (e) Transient reflectance trace at the center of WSe$_2$ excitonic transition cut of (c). (f) Second derivative of (d). All spectra were collected at sample temperature of 7 K.

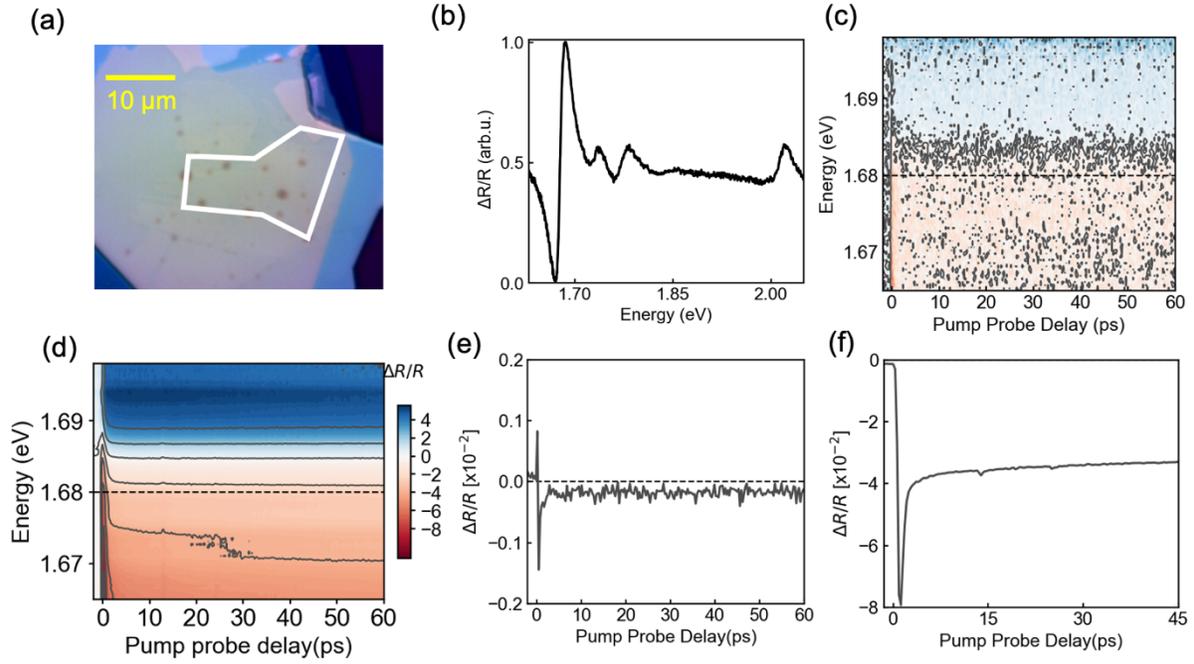

**Figure S4. Control WSe$_2$/WSe$_2$ device without graphite layers (D5): structure and characterization.** (a) Optical image of the device where the WSe$_2$/WS$_2$ overlap region are labeled with white line. (b) Steady-state reflectance spectrum of D5, showing the moiré WSe$_2$ transition and WS$_2$ transition. (c) Transient reflectance spectra plotted as a function of probe energy (covering the first moiré exciton of WSe$_2$). (d) Transient reflectance spectra plotted as a function of probe energy (covering the first moiré exciton of WSe$_2$) using the above-gap pumping condition (hν$_3$ = 3.1 eV). (e) Transient reflectance trace at the center of WSe$_2$ excitonic transition cut of fig c. (f) Transient reflectance trace at the center of WSe$_2$ excitonic transition cut of (d). No oscillation is observed in the control device without few-layer graphite gates, with below gap and above gap excitation. The pump fluence employed for the transient measurements was 13.7 μJ/cm$^2$. All spectra were collected at sample temperature of 7 K.

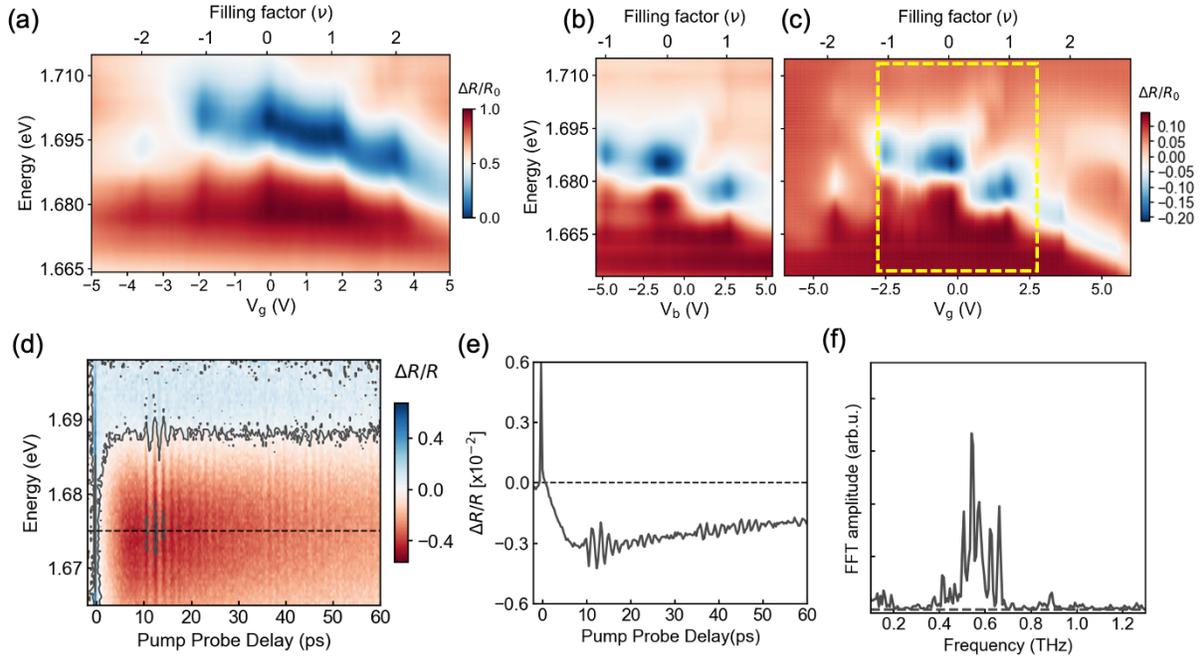

**Figure S5. Reflectance gate scan for D1 ($\theta = 60\pm1°$) and D3 ($\theta = 0\pm1°$) and $\nu = -2$ state of WSe$_2$/WS$_2$ device D1 ($\theta = 60\pm1°$):** *(a)* Gated ($V_g = V_t = V_b$) steady-state reflectance spectrum of the lowest energy moiré exciton of WSe$_2$ (T = 11 K). *(b)* Gated ($V_b$) steady-state reflectance of the lowest energy moiré band of WSe$_2$ (T = 8 K). (c) Symmetrically gated ($V_g = V_b = V_t$) steady-state reflectance spectrum prior to $V_t$ issues (T = 3 K). The yellow box highlights the accessible doping region via exclusively scanning $V_b$ (panel b, T = 8 K). A comparison shows no detectable change to the doping dependence of the device under a small static electric field. (d) Transient reflectance spectra plotted as a function of probe energy (covering the first moiré exciton of WSe$_2$) of $\nu = -2$ state in the WSe$_2$/WS$_2$ device D1. (e) Slice at a fixed probe energy corresponding to the dashed line in panel d. (f) Experimental FFT spectrum for the $\nu = -2$ state of D1 following population subtraction.

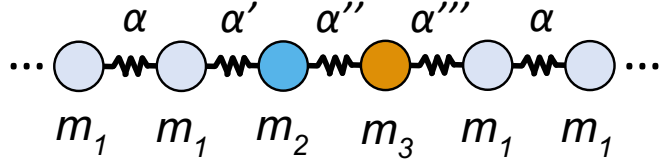

**Figure S6. Linear chain model schematic.** The TMD bilayer is represented by $m_2$ and $m_3$ and h-BN by $m_1$ with corresponding force constants.

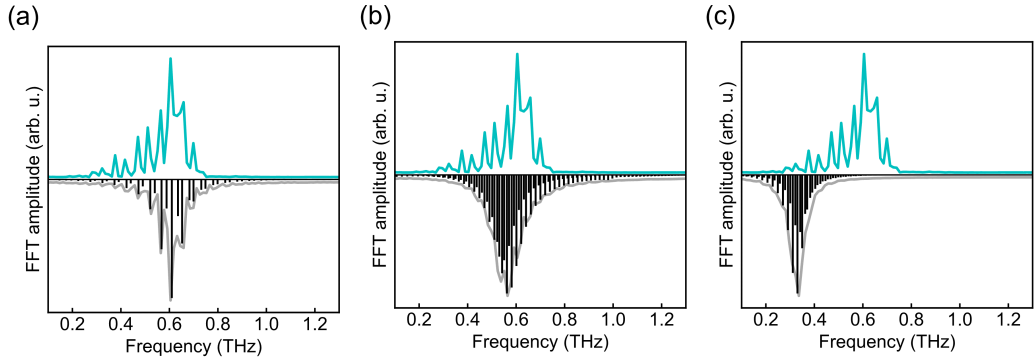

**Figure S7. D3 LCM prediction for LB versus S modes.** (a) LCM fit (shown as negative signal where the stick spectrum prediction is shown in black and broadened spectrum is shown in gray) for the LB modes compared to D3 experimental spectrum (positive signal shown in light blue). (b) LCM results for the S modes compared to experimental D3 spectrum where all parameters are fixed to those in Table S1 except $\alpha_{hBN\text{-}hBN}$ has been changed accordingly. (c) LCM results for the S modes compared to experimental D3 spectrum where $\alpha_{hBN\text{-}WSe2}$, $\alpha_{WSe2\text{-}WS2}$, and $\alpha_{hBN\text{-}WS2}$ have been scaled by 30% compared to the parameters in Table S1 and $\alpha_{hBN\text{-}hBN}$ has been changed accordingly. This scaling has been chosen based on the roughly expected decrease in the S versus LB force constant (Table 1 in Ref. (*7*)). It is not possible to obtain a reasonable fit with the LCM for the S modes based on the known ranges for $N_{hBN\,(top)}$ and $N_{hBN\,(bot.)}$.